\documentclass[fleqn,usenatbib]{mnras}

\usepackage{newtxtext,newtxmath}

\usepackage[T1]{fontenc}

\DeclareRobustCommand{\VAN}[3]{#2}
\let\VANthebibliography\thebibliography
\def\thebibliography{\DeclareRobustCommand{\VAN}[3]{##3}\VANthebibliography}

\usepackage{graphicx}	
\usepackage{xcolor, ulem}
\usepackage{subcaption}

\graphicspath{ {figures/} }

\title[Non-LTE double detonation simulations]{Non-LTE radiative transfer simulations: Improved agreement of the double detonation with normal Type Ia supernovae}

 \author[C. E. Collins et al.]{Christine E. Collins,$^{1}$\thanks{E-mail: ccollins@tcd.ie}
 Luke J. Shingles,$^{2}$
 Stuart A. Sim,$^{3}$ 
 Fionntan P. Callan,$^{3}$
 Sabrina~Gronow,$^{4,5}$
 \newauthor Wolfgang Hillebrandt,$^{6}$
 Markus Kromer,$^{5}$
 R\"{u}diger Pakmor$^{6}$
 and Friedrich~K.~R\"{o}pke$^{4,5,7}$
 \\
$^{1}$School of Physics, Trinity College Dublin, The University of Dublin, Dublin 2, Ireland\\
 $^{2}$GSI Helmholtzzentrum f\"{u}r Schwerionenforschung, Planckstraße 1, 64291 Darmstadt, Germany\\
 $^{3}$Astrophysics Research Centre, School of Mathematics and Physics, Queen's University Belfast, Belfast BT7 1NN, Northern Ireland, UK\\
 $^{4}$Zentrum f\"ur Astronomie der Universit\"at Heidelberg,
 Astronomisches Rechen-Institut, M\"{o}nchhofstr. 12-14, 69120 Heidelberg, Germany\\
 $^{5}$Heidelberger Institut f\"{u}r Theoretische Studien,
Schloss-Wolfsbrunnenweg 35, 69118 Heidelberg, Germany\\
$^{6}$Max-Planck-Institut f\"{u}r Astrophysik, Karl-Schwarzschild-Str. 1, D-85748, Garching, Germany\\
 $^{7}$Zentrum f\"ur Astronomie der Universit\"at Heidelberg, Institut f\"ur
Theoretische Astrophysik, Philosophenweg 12, 69120 Heidelberg, Germany\\
 }

\date{Accepted XXX. Received YYY; in original form ZZZ}

\pubyear{2024}

\begin{document}
\label{firstpage}
\pagerange{\pageref{firstpage}--\pageref{lastpage}}
\maketitle

\begin{abstract}
The double detonation is a widely discussed explosion mechanism for Type Ia supernovae, whereby a helium shell detonation ignites a secondary detonation in the carbon/oxygen core of a white dwarf.
Even for modern models that invoke relatively small He shell masses, many previous studies have found that the products of the helium shell detonation lead to discrepancies with normal Type Ia supernovae, such as strong \ion{Ti}{II} absorption features, extremely red light curves and too large a variation with viewing direction.
It has been suggested that non local thermodynamic equilibrium (non-LTE) effects may help to reduce these discrepancies with observations.
Here we carry out full non-LTE radiative transfer simulations for a recent double detonation model with a relatively small helium shell mass of 0.05 M$_\mathrm{\sun}$.
We construct 1D models representative of directions in a 3D explosion model to give an indication of viewing angle dependence.
The full non-LTE treatment leads to improved agreement between the models and observations.
The light curves become less red, due to reduced absorption by the helium shell detonation products, since these species are more highly ionised.
Additionally, the expected variation with observer direction is reduced.
The full non-LTE treatment shows promising improvements, and reduces the discrepancies between the double detonation models and observations of normal Type Ia supernovae.

\end{abstract}




\section{Introduction}

Sub-Chandrasekhar mass (Sub-M$_{\rm Ch}$) white dwarfs are promising progenitors for Type Ia supernovae (SNe~Ia) \citep{shigeyama1992a, nugent1997a, hoeflich1998a,
    sim2010a, blondin2017a, shen2018b}.
The detonations of bare sub-M$_{\rm Ch}$ carbon-oxygen (C/O) white dwarfs are able to produce a reasonable number of
observed features of SNe Ia \citep{sim2010a, blondin2017a, shen2018b},
including accounting for a range of brightnesses by varying white dwarf mass.
They show light curves with rise times
and peak colours similar to normal SNe Ia and can roughly reproduce the overall trend of the width-luminosity
relationship \citep{phillips1993a}.
However, the physical mechanism igniting the detonation
in these bare C/O white dwarf models was not considered.
The double detonation \citep[see e.g.,][]{taam1980a, nomoto1980a, nomoto1982a, livne1990a, woosley1994b,
    hoeflich1996a, nugent1997a} is a widely discussed mechanism
to ignite a sub-M$_{\rm Ch}$ white dwarf.
In this scenario, a helium detonation is ignited in a surface helium shell on a
sub-M$_{\rm Ch}$ C/O white dwarf.
The helium detonation then ignites a secondary carbon detonation in the
core.
Early versions of double detonation models
invoked relatively massive helium shells of $\sim 0.2$~M$_\odot$
\citep{nugent1997a},
and predicted light curves and spectra that were not consistent with those of normal SNe~Ia,
due to over-production of iron group elements.
Recent work has shown that considerably lower mass helium shells ($< 0.1$ M$_\odot$) may be able to ignite
a secondary core detonation \citep{bildsten2007a, shen2009a, fink2010a, shen2010a}.
This has contributed to renewed interest in the
double detonation explosion scenario,
since this reduces the discrepancies with observations caused by the over production
of iron group elements at high velocities, produced in the helium
shell detonation \citep{kromer2010a, woosley2011b}.
However, compared to the detonation of bare C/O sub-M$_{\rm Ch}$ white dwarfs,
the detonation of the helium shell introduces discrepancies with observations,
such as strong absorption features (particularly by \ion{Ti}{II}) and red light curve colours.

The greatest source of discrepancy for the double detonation scenario
compared to observations, is due to the burning products of the He shell detonation.
Large abundances of heavy elements are synthesised in the outer layers of
the ejecta during the He detonation, and the burning products
(predominantly Cr and Ti, and Fe-peak elements)
cause line blanketing of blue wavelengths
that produce strong absorption features not observed in normal
SNe~Ia \citep{kromer2010a, woosley2011b, sim2012a, polin2019a, gronow2020a, gronow2021a, shen2021b, collins2022b}.
Recent simulations considering a minimal helium shell mass (around 0.02~M$_\odot$)
have shown that double detonations may be able to reproduce normal SNe Ia \citep{townsley2019a, shen2021b}, although \citet{collins2022b} showed that a minimal He shell mass does not necessarily lead to spectra and light curves in agreement with normal SNe Ia.
These minimal mass He shell simulations considered a traditional double detonation, where He is accreted slowly and the He shell becomes unstable to detonation. 
In dynamical accretion scenarios \citep[e.g.,][]{guillochon2010a, dan2011a, raskin2012a, pakmor2013a, shen2018a, pakmor2022a} the minimum He shell mass required to reach ignition is smaller still.

Many radiative transfer simulations for the recent generation of lower
He shell mass double detonation models have assumed
local thermodynamic equilibrium (LTE) \citep[e.g.][]{woosley2011b, polin2019a, townsley2019a, shen2021b},
or used approximations for non-LTE \citep{kromer2010a, gronow2020a, collins2022b}.
\citet{dessart2014b} discuss the importance of non-LTE effects for SNe Ia modelling.
\citet{nugent1997a} carried out non-LTE radiative transfer simulations
for early, thick He shell double detonation models, but these models were not able to reproduce observations of normal SNe~Ia.
\citet{shen2021a} carried out non-LTE simulations for models of bare C/O white dwarf explosions with no He shell detonation and found improved agreement between the models and the Phillips relation \citep{phillips1993a}.
Recently, \citet{boos2024a} presented 1D non-LTE simulations based on different lines of sight in a 2D double detonation model which considered a minimal He shell mass (0.016 M$_\odot$).
This led to bluer colours and slower declines from maximum, generally showing greater agreement with observations,
although for some models the colours appear too blue and declines too slow. 
In this paper we carry out full non-LTE simulations for a recent
low He shell mass model (with a modest helium shell mass of 0.05~M$_\odot$).

\section{Methods}

\subsection{Radiative transfer}
We use the time-dependent multi-dimensional Monte Carlo radiative transfer
code \textsc{artis} (\citealt{sim2007b, kromer2009a}, based on the
methods of \citealt{lucy2002a, lucy2003a, lucy2005a}) to calculate
synthetic light curves and spectra.
\textsc{artis} has been extended by \citet{Shingles2020a} to include 
a full solution to the non-LTE equations of statistical equilibrium
for atomic level populations, a non-thermal solver, detailed bound-free estimators, and a non-LTE radiation field model.
We refer to this version of \textsc{artis} by \citet{Shingles2020a} as \mbox{\textsc{artis-nlte}}.
We compare simulations using \textsc{artis-nlte} to simulations using the approximate
non-LTE treatment in \textsc{artis}, as described by \citet{kromer2009a},
which we now refer to as \textsc{artis-classic}.
Specifically we use the ``detailed ionisation'' treatment, which uses a nebular approximation to parametrise the radiation field with a dilute blackbody at the radiation temperature (T$_\mathrm{R}$).
Level populations are solved using the Boltzmann formula evaluated at a temperature corresponding to the local energy density of the radiation field (T$_\mathrm{J}$). See \citet{kromer2009a} for details of this treatment.
While intended to capture some non-LTE behaviour, this treatment is still based on assumptions made under LTE (e.g. that the level populations can be described by the Boltzmann formula).

The \textsc{artis-nlte} simulations are carried out in the same way as the non-LTE simulation presented by \citet{collins2023b}. 
In this paper we present 1D simulations.
Future work will be to carry out multi-dimensional full non-LTE
simulations.

For the \textsc{artis-nlte} calculations, we use the same atomic dataset as \citet{collins2023b},
based on the atomic data compilation of
\textsc{cmfgen}\footnote{Available at \url{http://kookaburra.phyast.pitt.edu/hillier/web/CMFGEN.htm}}
(\citealt{hillier1990a, hillier1998a}), similar to the dataset compiled by \citealt{Shingles2020a}).
The atomic data used in the \textsc{artis-classic} simulations is the `big-gf-4' atomic data set
of \cite[][see their table 1]{kromer2009a} and \citet{gall2012a}, sourced from \citet{kurucz1995a} and \citet{kurucz2006a}.

\subsection{Models}
\label{sec:1Dmodels}

In this work we carry out 1D simulations,
however, since the double detonation scenario is highly asymmetric
\citep[e.g.,][]{kromer2009a, gronow2020a, gronow2021a, boos2021a},
we would like to determine the potential effects of
non-LTE on the viewing angle dependence of the synthetic observables.
Therefore, 
we create 1D models from ejecta in different directions from a 3D explosion model (described in Appendix~\ref{sec:verify1Dmodels}).
We chose to investigate Model M2a from \citet{gronow2020a}, using the same methods to construct 1D models as \citet{collins2023b}.
Model M2a had a total mass of 1.05 M$_\odot$, with a relatively low helium shell mass of 0.05 M$_\odot$.

\begin{figure*}

\includegraphics[width=1.\textwidth]{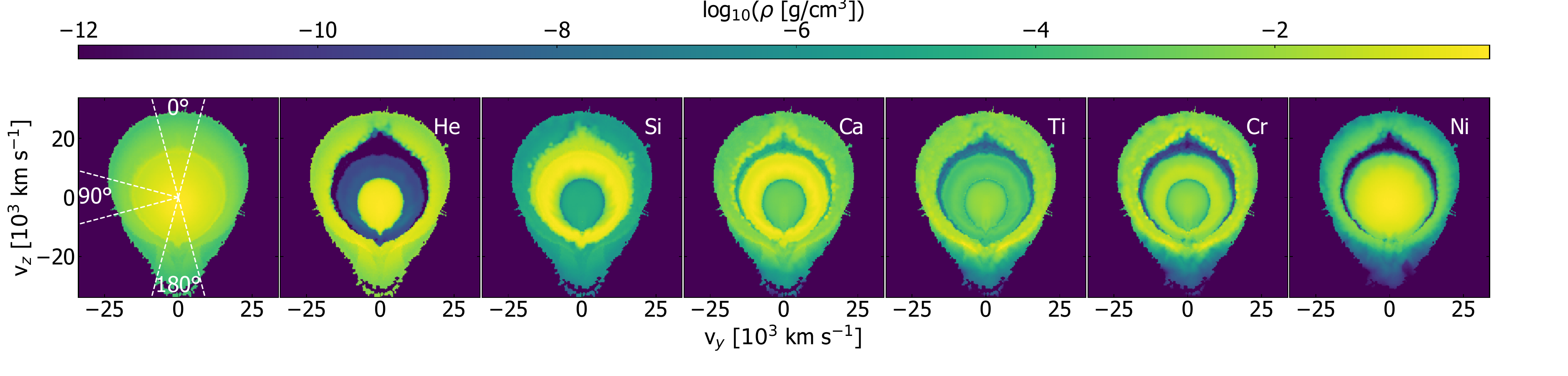}

\caption{Density (left panel) of the 3D Model M2a and the density of key elements at 100 s after explosion in a slice through the $x$-axis ($x=0$), indicating the model ejecta structure.
The dashed lines in the left panel indicate the ejecta lying within the cones used to create the 1D profiles of different directions within the model.}
\label{fig:3Dmassfractions}
\end{figure*}

\begin{figure}
\includegraphics[width=0.45\textwidth]{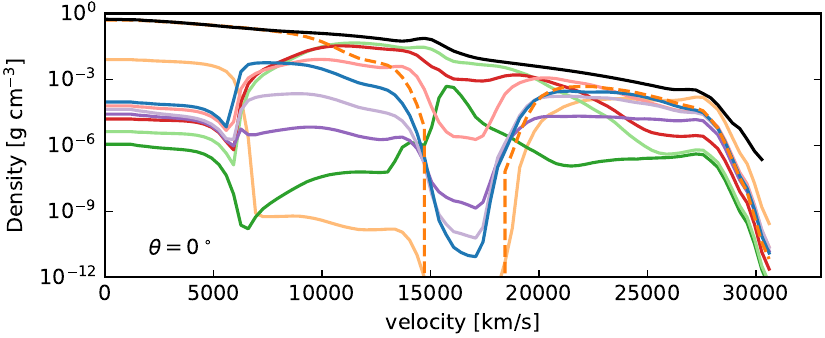}

\includegraphics[width=0.45\textwidth]{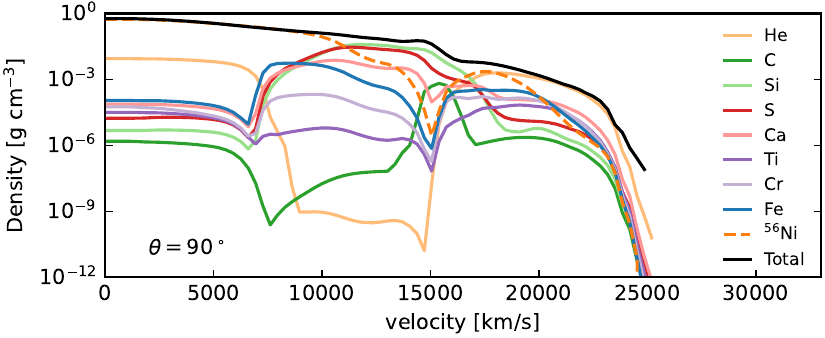}

\includegraphics[width=0.45\textwidth]{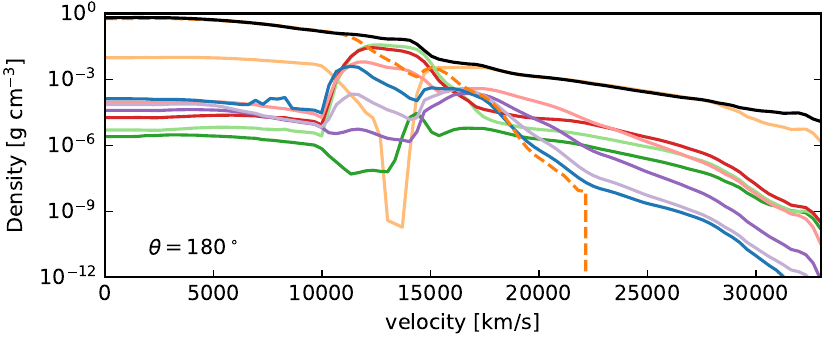}

\caption{Density profiles of the 1D models representing directions in a 3D double detonation model at 100 s after explosion.}
\label{fig:abundances1D}
\end{figure}

The structure of the ejecta in the 3D Model M2a is indicated by Figure~\ref{fig:3Dmassfractions}.
We investigate three directions.
Firstly, viewing towards the initial ignition point at the pole on the $+z$ axis ($\theta = 0^\circ$),
where the highest abundance of intermediate mass elements (IMEs), and the
lowest mass of $^{56}$Ni
were synthesised in Model M2a.
Secondly, an equatorial line of sight ($\theta = 90^\circ$), which gives similar results to the
angle averaged values found for Model M2a by \citet{gronow2020a}.
This is the 1D model that \citet{collins2023b} considered.
The third line of sight chosen is the antipode of the initial ignition
point ($\theta = 180^\circ$) where the helium detonation converged.
This line of sight has
the most extended layer of unburnt helium and the lowest abundance
of IMEs.
It also has the highest abundance of $^{56}$Ni from the core detonation, since the core detonation was ignited in this direction
(see Table \ref{tab:initialabundaces1Dslices}).
In Appendix \ref{sec:verify1Dmodels} we test the agreement between
the 1D models and the line of sight
each model represents in the 3D simulation of model M2a (using \textsc{artis-classic}).
The 1D models show reasonable agreement with the 3D model until around maximum light. After maximum light, as the ejecta become less optically thick, the 1D models overestimate the extent of the viewing angle dependence (see Figure~\ref{fig:lightcurves1Dcomparedto3D}).
We therefore limit our radiative transfer simulations here to early times.
The need for multi-dimensional radiative transfer simulations has previously been discussed (e.g., see \citealt{pakmor2024a}).
The density profiles of the three 1D models representing the chosen lines of sight are shown in
Figure~\ref{fig:abundances1D}.

\begin{table}
\centering
\caption{Masses of select species in the 1D models representative of a viewing angle of 
$\theta = 0^\circ$, $90^\circ$ and $180^\circ$ of the 3D Model M2a.}
\label{tab:initialabundaces1Dslices}
\begin{tabular}{llll}
\hline
Species     & $\theta = 0^\circ$ & $\theta = 90^\circ$ & $\theta = 180^\circ$ \\
     & {[}M$_{\odot}${]}  & {[}M$_{\odot}${]}   & {[}M$_{\odot}${]}    \\ 
     \hline
He   & 1.115$\times 10 ^{-2}$          & 1.846$\times 10 ^{-2}$           & 4.554$\times 10 ^{-2}$            \\
C    & 4.872$\times 10 ^{-4}$          & 1.062$\times 10 ^{-3}$           & 7.690$\times 10 ^{-5}$            \\
O    & 6.164$\times 10 ^{-2}$          & 4.743$\times 10 ^{-2}$           & 1.192$\times 10 ^{-2}$            \\
Mg   & 1.647$\times 10 ^{-2}$          & 1.066$\times 10 ^{-2}$           & 2.255$\times 10 ^{-3}$            \\
Si   & 1.899$\times 10 ^{-1}$          & 1.642$\times 10 ^{-1}$           & 9.023$\times 10 ^{-2}$            \\
S    & 1.166$\times 10 ^{-1}$          & 1.084$\times 10 ^{-1}$           & 6.464$\times 10 ^{-2}$            \\
Ca   & 2.863$\times 10 ^{-2}$          & 2.538$\times 10 ^{-2}$           & 1.601$\times 10 ^{-2}$            \\
Ti   & 6.673$\times 10 ^{-4}$          & 6.178$\times 10 ^{-4}$           & 1.266$\times 10 ^{-3}$            \\
Cr   & 2.663$\times 10 ^{-3}$          & 1.908$\times 10 ^{-3}$           & 2.364$\times 10 ^{-3}$            \\
Co   & 4.011$\times 10 ^{-4}$          & 4.292$\times 10 ^{-4}$           & 6.214$\times 10 ^{-4}$            \\
$^{56 }$Ni & 3.202$\times 10 ^{-1}$          & 4.880$\times 10 ^{-1}$           & 6.692$\times 10 ^{-1}$            \\ \hline           
\end{tabular}
\end{table}

\section{Results}

In this section we present the synthetic observables calculated for our 1D models using \textsc{artis-nlte},
and compare these to the \textsc{artis-classic} simulations.
All simulations presented in this section are for the 1D models described in Section~\ref{sec:1Dmodels}.

\subsection{Non-LTE light curves}
\label{sec:lightcurvesNLTEdoubledet}

\begin{figure*}
\includegraphics[width=0.95\textwidth]{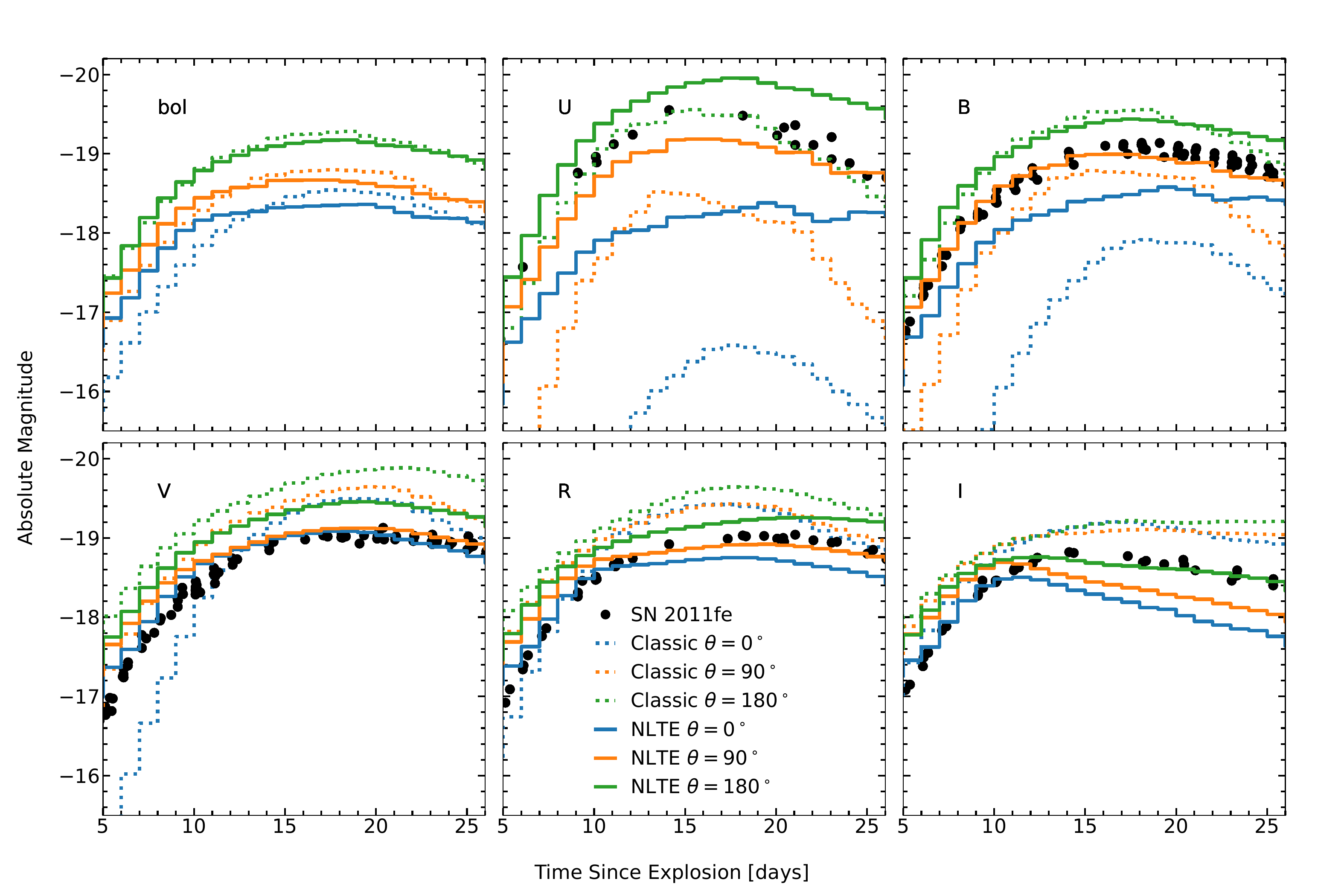}

\caption{Light curves for the 1D models.
The dotted lines show the light curves that \textsc{artis-classic} predicts,
and the solid lines show the \textsc{artis-nlte} light curves.
The light curves of SN 2011fe are also plotted for reference.
}
\label{fig:lightcurves_1DdoubledetNLTE}
\end{figure*}

We show the light curves for our models in
Figure~\ref{fig:lightcurves_1DdoubledetNLTE} and compare the \textsc{artis-nlte} simulations to the \textsc{artis-classic} simulations.
To reduce the level of Monte Carlo noise in the light curves of the \textsc{artis-nlte} simulations, we have binned the emitted flux into 1 day intervals.
We use the same binning for the \textsc{artis-classic} simulations to provide a direct comparison.
The \textsc{artis-nlte} simulations show significant changes
compared to the \textsc{artis-classic} simulations.

\subsubsection{Non-LTE double detonation colour evolution}
\label{sec:colourNLTEdoubledet}

\begin{figure*}
\includegraphics[width=0.85\textwidth]{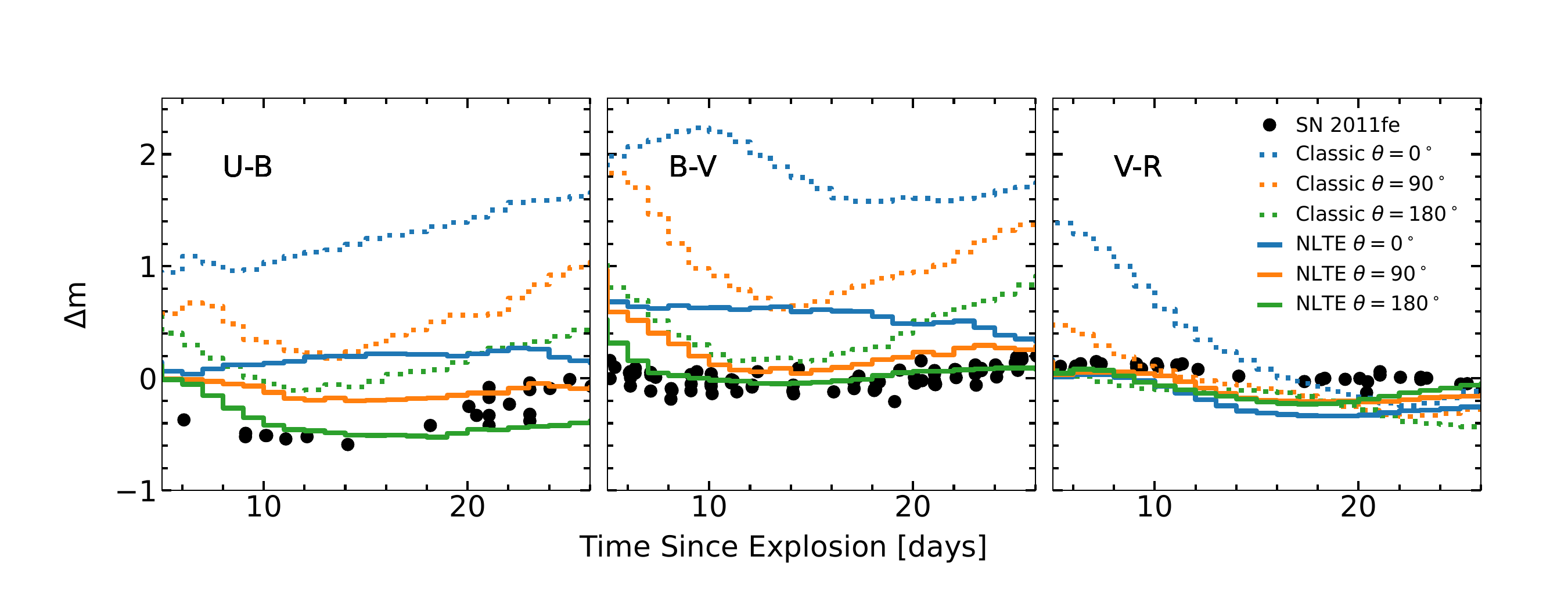}

\caption{Model colours for the 1D models representative of lines of sight in
Model M2a, calculated using \textsc{artis-nlte} (solid lines) and \textsc{artis-classic}
(dotted lines).
We also plot the colours of the normal SN~2011fe for comparison.
}
\label{fig:colorevolution_1DdoubledetNLTE}
\end{figure*}

Double detonation simulations tend to show colours too red compared to normal SNe Ia, due to absorption and line blanketing at blue wavelengths by the He shell detonation products (e.g., Ti and Cr).
As can be seen in Figure~\ref{fig:lightcurves_1DdoubledetNLTE}, the \textsc{artis-classic} simulations show U and B band light curves that are significantly fainter than the redder bands, as can be seen from the colour evolution of these models in Figure~\ref{fig:colorevolution_1DdoubledetNLTE}.
In particular the models representing the fainter directions at $\theta = 0^\circ$ and $\theta = 90^\circ$ show extremely red colours.

The simulations using \textsc{artis-nlte} are brighter in the U and B bands compared to the \textsc{artis-classic} simulations, particularly the models at $\theta = 0^\circ$ and at $\theta = 90^\circ$. Additionally, the \textsc{artis-nlte} simulations are fainter in V band. As such, the colours of the \textsc{artis-nlte} simulations are significantly less red than the \textsc{artis-classic} simulations, as can be seen in Figure~\ref{fig:colorevolution_1DdoubledetNLTE}.
The bluer colours in the \textsc{artis-nlte} simulations are due to the reduced absorption at blue wavelengths, which results from the non-LTE treatment. Specifically, as a result of the higher ionisation states predicted by the \textsc{artis-nlte} simulations (discussed further in Section~\ref{sec:ionisationstate}).
More flux escapes in the U and B bands, which is not absorbed and remitted through fluorescence in the V band, as was predicted by the \textsc{artis-classic} simulations.
This demonstrates that the colour from double detonation simulations is strongly dependent on the assumptions made in the radiative transfer.

\citet{boos2024a} also found this trend of bluer colours for their non-LTE simulations, however, we note that the minimal He shell mass model they considered did not show colours as extremely red in LTE as the model we consider here, owing to lower masses of heavy elements produced in their He shell detonation. For example, the reddest line of sight in their 2D LTE simulation showed a \mbox{B-V} colour at B max of (B-V)$_\mathrm{Bmax} \approx 0.26$, whereas for our reddest \textsc{artis-classic} simulation (B-V)$_\mathrm{Bmax} =1.6$.

\subsubsection{Direction dependence}

As discussed by \citet{gronow2020a} for the 3D Model M2a, the direction at $\theta = 0^\circ$ is the faintest, while the direction at $\theta = 180^\circ$ is the brightest.
In particular, the U and B bands showed an extreme variation with observer viewing angle, that appears to be too strong compared to observations of SNe~Ia \citep{collins2022b}.
The \textsc{artis-classic} 1D simulations are similar to those
presented by \citet{gronow2020a} for the 3D Model M2a in specific lines of sight (see Appendix~\ref{sec:verify1Dmodels}), and therefore are representative of the angle variation in the 3D model.

The viewing angle dependence indicated by the 1D models at each orientation in the U and B bands for the
\textsc{artis-nlte} simulations
is significantly less compared to the \textsc{artis-classic} simulations. 
In B band, the \textsc{artis-nlte} simulations vary by $\sim 1$ mag, compared to $\sim 2$~mag for the \textsc{artis-classic} simulations. In U band, the difference is even more extreme, as the \textsc{artis-classic} simulations vary by $\sim 3$~mag, while the \textsc{artis-nlte} simulations vary by $\sim 1.5$ mag.
The difference between the \textsc{artis-classic} and \textsc{artis-nlte} simulations is greatest for the model at $\theta = 0^\circ$, but is also significant for the model at
$\theta = 90^\circ$.

In the \textsc{artis-classic} simulations, these models
showed 
red colours
due to burning products of the helium shell detonation,
in particular by \ion{Ti}{II} and \ion{Cr}{II}.
Since the amount of He detonation ash varies with orientation (see Figure~\ref{fig:abundances1D}),
the degree of redness shown by the light curves also varies significantly with observer orientation.
This leads to the large viewing angle dependence in the U and B bands.

The reduced absorption at blue wavelengths in the \textsc{artis-nlte} simulations leads to much less variation than was found using \textsc{artis-classic} in the U and B bands, indicating a significantly reduced observer angle dependence.
The reduced absorption is due to the higher ionisation state predicted by \textsc{artis-nlte} (discussed further in Section~\ref{sec:spectra_non-LTEdoubledet}).
Thus, using a full non-LTE treatment for the radiative transfer appears to reduce the viewing angle dependence shown by double detonation simulations.
In future, multi-dimensional non-LTE simulations should verify this.

We note that the difference in the bolometric light curves between the \textsc{artis-nlte} and \textsc{artis-classic} simulations is much less significant than for the band-limited light curves. This indicates that bolometric light curves are less sensitive to non-LTE effects (as has previously been suggested, e.g., by \citealt{wygoda2019a}), although the changing mean opacity does affect even the bolometric curve, particularly for the fainter model.

\subsubsection{Comparison to normal SNe Ia}

In Figures~\ref{fig:lightcurves_1DdoubledetNLTE} and \ref{fig:colorevolution_1DdoubledetNLTE} we also plot the observations of SN~2011fe \citep{nugent2011a},
which is an extremely well observed normal SNe Ia,
and is representative of light curves typical of
normal SNe Ia.

The \textsc{artis-classic} simulation at $\theta = 0^\circ$ is significantly too red compared to SN~2011fe, and the \textsc{artis-classic} simulation at $\theta = 90^\circ$ is also too red, as was discussed for the 3D model M2a by \citet{gronow2020a}.
While the \textsc{artis-classic} light curves at $\theta = 180^\circ$ are brighter than SN~2011fe in most bands, the colours are the most similar to those of SN~2011fe.
The \textsc{artis-nlte} simulations show colours much more similar to SN~2011fe for all models.
The \textsc{artis-nlte} simulation at $\theta = 0^\circ$ is still $\sim 0.5$ mag redder than SN~2011fe in U-B and \mbox{B-V} colour, but it is no longer as extremely red as the \textsc{artis-classic} simulation predicts. 

The bluer colours resulting from the \textsc{artis-nlte} simulations improves the agreement with normal SNe Ia, and in particular, the model at $\theta = 90^\circ$ shows good agreement with the light curves and colour evolution shown by SN~2011fe.
The model at $\theta = 90^\circ$ is representative of an equatorial line of sight, and therefore represents the most probable observed direction (if drawn from a random distribution of solid angle).

The full non-LTE simulations show that double
detonations are able to reproduce normal SNe Ia light curves,
without necessarily invoking a minimal He shell mass \citep[e.g., the models by][]{townsley2019a, shen2021b}.
In non-LTE, the heavy elements produced in the helium shell detonation can be hidden, due to the higher ionisation states.
In future, this should be confirmed with multi-dimensional simulations, since the double detonation is highly asymmetric.

\subsection{Non-LTE spectra}
\label{sec:spectra_non-LTEdoubledet}

\begin{figure*}
    \includegraphics[width=0.45\textwidth]{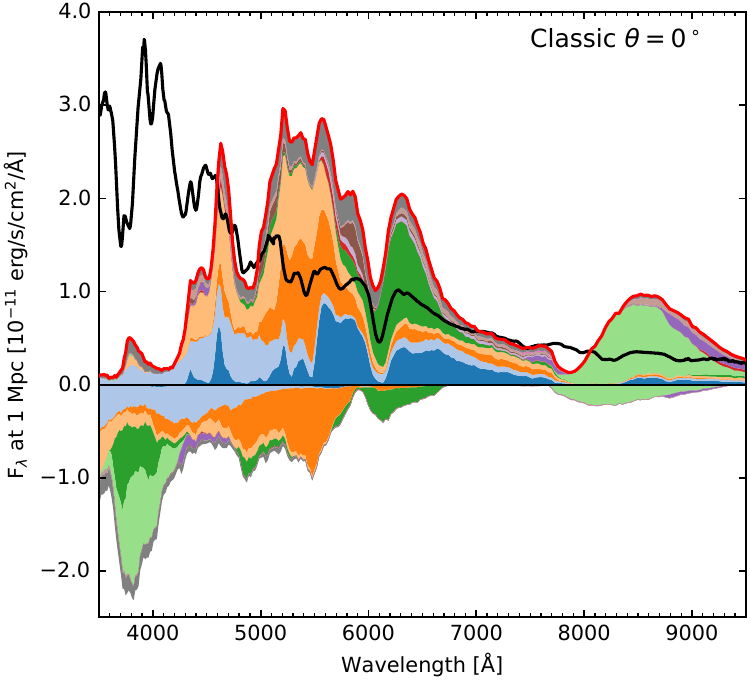}\includegraphics[width=0.45\textwidth]{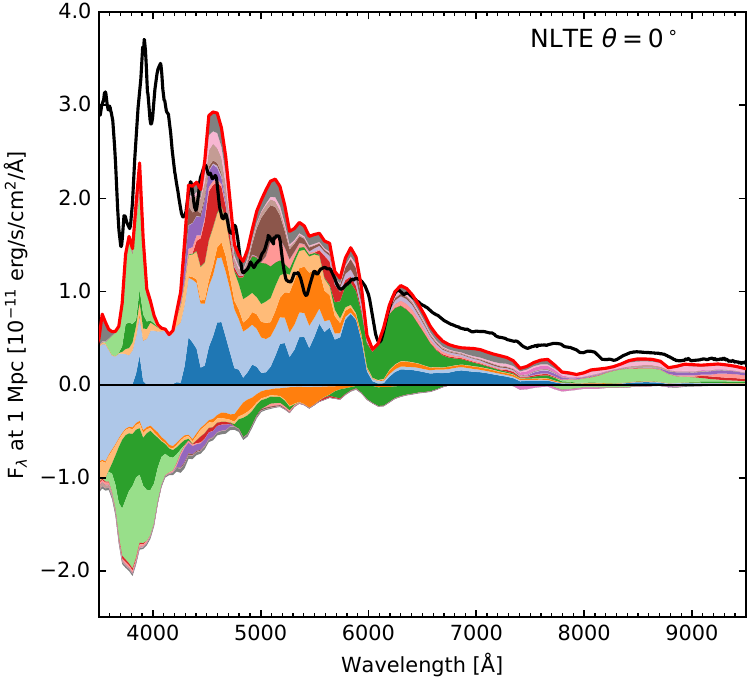}
    \includegraphics[width=0.45\textwidth]{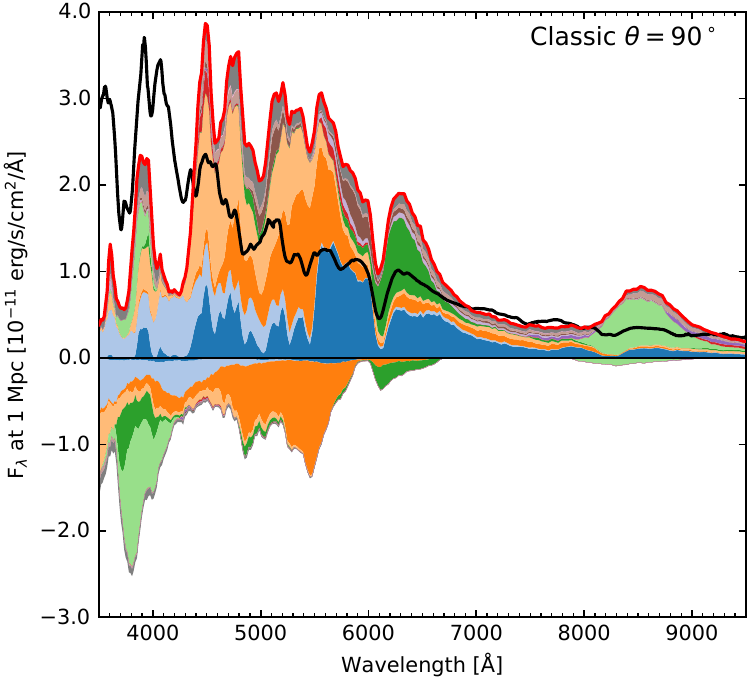}\includegraphics[width=0.45\textwidth]{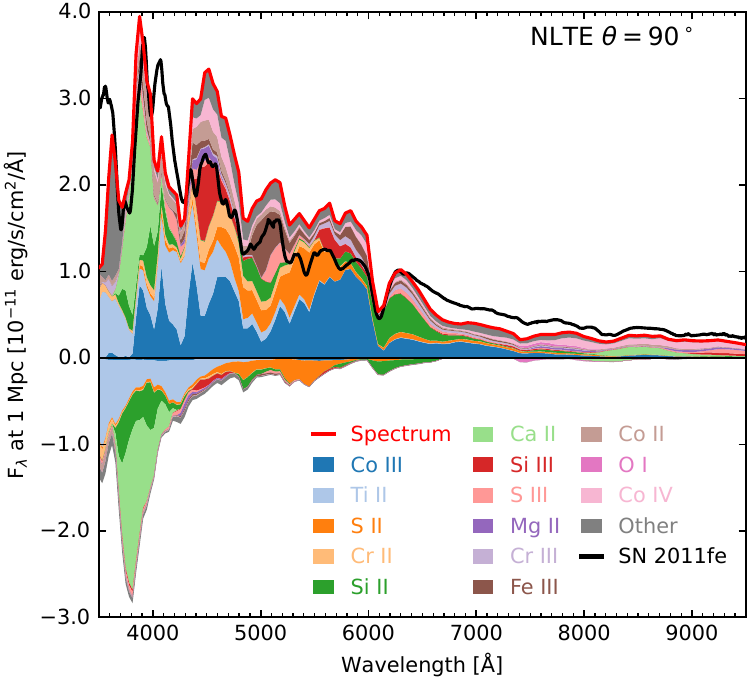}
    \includegraphics[width=0.45\textwidth]{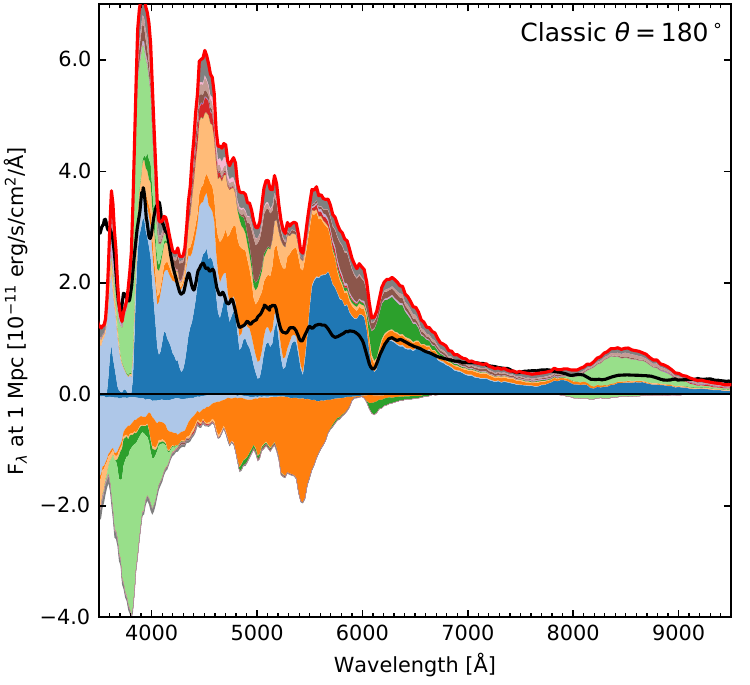}\includegraphics[width=0.45\textwidth]{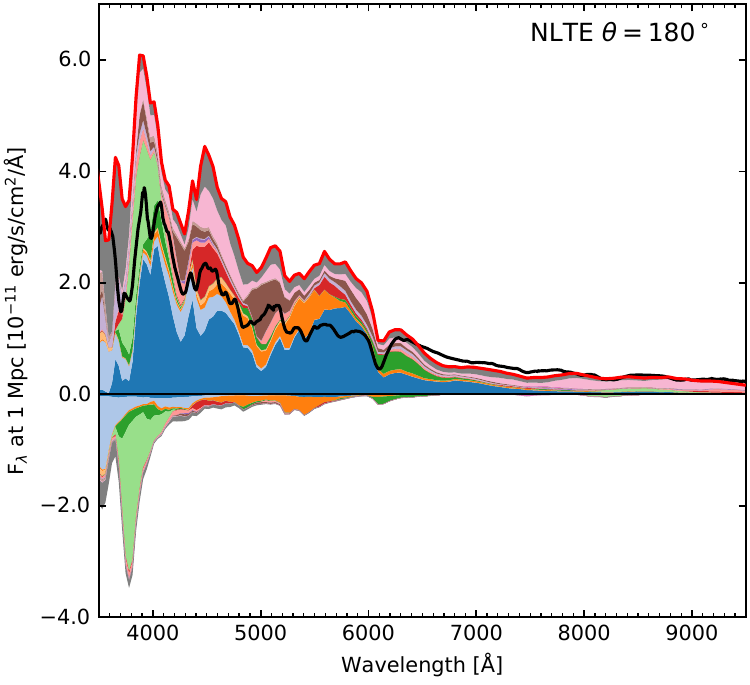}

    \caption{
    Model spectra at 18 days after explosion. The left hand panels show the \textsc{artis-classic} simulations while the right hand panels show the \textsc{artis-nlte} simulations.
    The colours indicate the relative contributions of select species to shaping the model spectra. 
    Absorption processes are indicated below the axis.
    In each panel, a spectrum of SN~2011fe near maximum light is plotted for comparison.
    }
    \label{fig:specemission_1DdoubledetNLTE}
\end{figure*}

We now discuss the changes to the spectra, and the contributions of
ions responsible for shaping the spectra.

In Figure \ref{fig:specemission_1DdoubledetNLTE}
we show the contributions to the spectra
at 18 days after explosion for
the \textsc{artis-nlte} and \textsc{artis-classic} simulations.
In the radiative transfer simulations,
we record details of the last interaction each Monte Carlo
packet underwent before escaping the ejecta.
For each wavelength bin in the synthetic spectrum, the area under
the spectrum is colour coded in proportion to the energy
carried by packets in that wavelength bin whose last
interaction was with each of the ions considered.
We also construct an equivalent histogram based
on where the wavelength bin packets were prior to their last
interaction (i.e.\ indicating where packets last underwent
absorption/scattering/fluorescence) and plot this
on the negative axis under the spectrum to
indicate the key absorption processes.

By comparing the model spectra in
Figure~\ref{fig:specemission_1DdoubledetNLTE}
the impact of the full non-LTE treatment in \textsc{artis-nlte}
can clearly be seen.
The greatest differences between the \textsc{artis-nlte}
and \textsc{artis-classic} spectra are the contributions from singly ionised elements.
The \textsc{artis-nlte} simulations are more highly ionised than the \textsc{artis-classic} simulations.
In particular, the \textsc{artis-nlte} simulations show less \ion{Ca}{II} triplet emission (with wavelengths of 8498, 8542 and 8662 \AA)
at this time and
less overall contribution from \ion{S}{II}, while
the \ion{S}{II} $\lambda \lambda$5454, 5640 `W' feature
is much more clearly defined.
We also point out that the strength of the \ion{Si}{II} feature at 6355 \AA \ changes.

The absorption and line blanketing due to \ion{Ti}{II} and \ion{Cr}{II} is significantly reduced.
\ion{Cr}{II} and \ion{Ti}{II} contributed to the redness
of the \textsc{artis-classic} simulations, in particular
for the $\theta = 0^\circ$ and $\theta = 90^\circ$ models.
While the \textsc{artis-nlte} simulations also show \ion{Ti}{II}
absorption, the line blanketing and strong absorption
is no longer as extreme as was found for the
\textsc{artis-classic} simulations, as in the \textsc{artis-nlte}
simulations there is less \ion{Cr}{II} absorption blended with the
\ion{Ti}{II}.

We note that in the \textsc{artis-nlte} simulations, contributions from doubly ionised species appear that were not present in the \textsc{artis-classic} simulations. For example, \ion{Si}{III} and \ion{Fe}{III} (see Figure~\ref{fig:specemission_1DdoubledetNLTE}).
This is also a result of the higher ionisation states predicted by the \textsc{artis-nlte} simulations.
\citet{boos2024a} also found a \ion{Si}{III} $\lambda4560$ feature emerged in their non-LTE simulations.

\subsubsection{Comparison to observations}

Also shown in Figure~\ref{fig:specemission_1DdoubledetNLTE} is the comparison to SN~2011fe. 
We find that the \textsc{artis-nlte}
simulations show improved overall agreement with the spectral energy distribution, 
as well as better agreement with individual spectral features in SN~2011fe.
For example, in the \textsc{artis-nlte} simulations, the strength of the \ion{Ca}{II} triplet emission is no longer
significantly over-predicted as it was by the \textsc{artis-classic} simulations, and the strength of the characteristic \ion{Si}{II} 6355 \AA \ feature is better matched (in particular for the most probable observer orientation at $\theta = 90^\circ$).

At $\theta=0^\circ$ in the \textsc{artis-nlte} simulation the \ion{Ti}{II} absorption trough around $\sim 4150$ \AA \ remains too strong compared to SN~2011fe, although we note that such a feature has been observed in peculiar SNe~Ia suggested to be the result of double detonations (e.g. SN~2016jhr; \citealt{Jiang2017a}). 
A \ion{Ti}{II} absorption feature also forms in the \textsc{artis-nlte} simulation at $\theta = 90^\circ$, which still appears slightly too strong compared to SN~2011fe, however, the agreement is significantly improved compared to the \textsc{artis-classic} simulation.

\subsubsection{Helium spectral features}

\begin{figure}
    \includegraphics[width=0.45\textwidth]{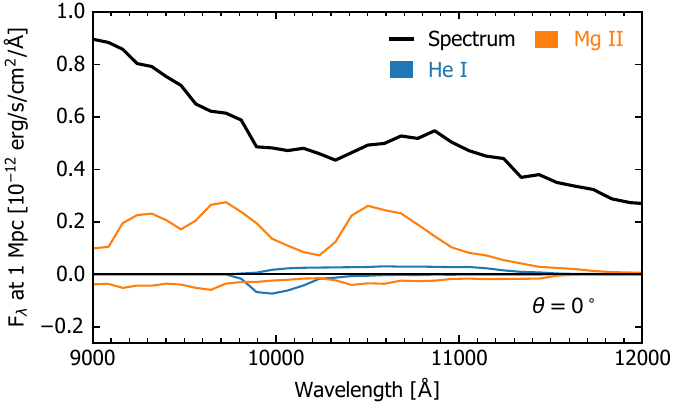}
    \includegraphics[width=0.45\textwidth]{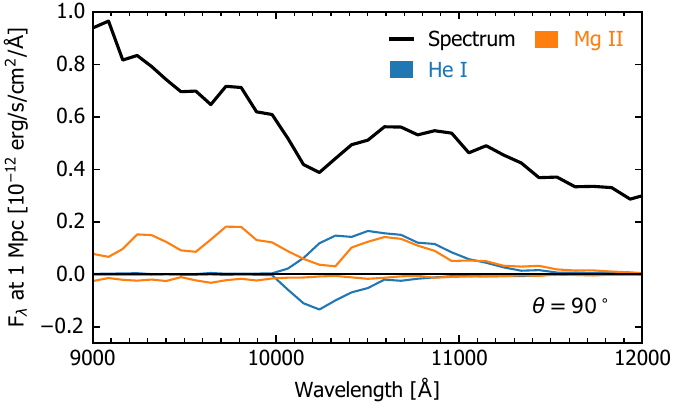}
    \includegraphics[width=0.45\textwidth]{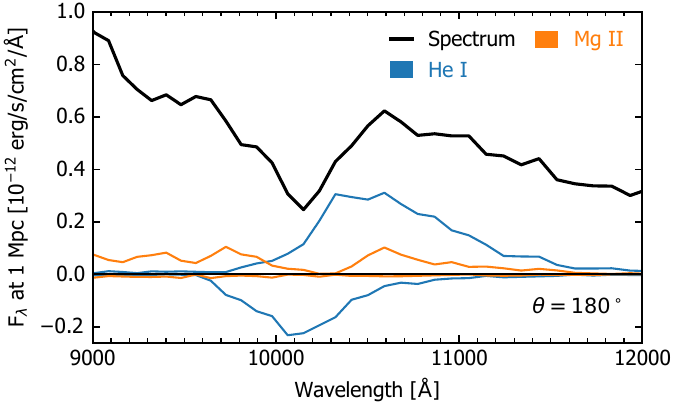}

    \caption{Spectra from the \textsc{artis-nlte} simulations at 5 days after explosion showing that a \ion{He}{I} 10830 \AA \ feature is predicted to form for all 1D models at this time.
    The net spectrum is shown as well as the relative contributions from \ion{He}{I} and \ion{Mg}{II}.
    In each panel, the absorption due to \ion{He}{I} is indicated beneath the axis (blue line), and corresponds to the absorption feature in the net spectrum at the same wavelength.
    The strength of the predicted \ion{He}{I} absorption feature is model dependent, and increases with increasing masses of He and $^{56}$Ni.
    No \ion{He}{I} feature is predicted by the \textsc{artis-classic} simulations.
    We note that we do not find a feature from \ion{C}{I} 10693 \AA \ in these models.
    }
    \label{fig:spectraHe}
\end{figure}

We also note that in the \textsc{artis-nlte} simulations a \ion{He}{I} 10830 \AA \ feature is predicted to form at early times for all three 1D models.
This is shown in Figure~\ref{fig:spectraHe} where we indicate the contributions from \ion{He}{I} and \ion{Mg}{II} in a similar way to Figure~\ref{fig:specemission_1DdoubledetNLTE}. The area enclosed by the coloured lines shows the proportion of energy emitted (above the axis) or absorbed (beneath the axis) where the last interaction of the Monte Carlo packet was with either \ion{He}{I} or \ion{Mg}{II}.

In the model at $\theta=0^\circ$, a \ion{He}{I} 10830~\AA \ absorption feature forms to the blue wing of the \ion{Mg}{II} 10927~\AA \ feature already at 5 days after explosion, and fades quickly.
As discussed by \citet{collins2023b},
the \ion{He}{I} feature in the model at $\theta=90^\circ$ is initially blended with the \ion{Mg}{II} feature and then separates to form at the blue wing with time.
In the model at $\theta=180^\circ$, the \ion{He}{I} feature is the strongest of the three models, and dominates the \ion{Mg}{II} feature. This 1D model has the highest mass of unburnt He and of $^{56}$Ni.
In this model, the feature remains blended with the \ion{Mg}{II} feature over time, and contributes until around 20 days after explosion.

The \ion{He}{I} feature is not present in the \textsc{artis-classic} simulations, since non-thermal energy is required to sufficiently excite helium to form a spectral feature.
\citet{callan2024a} have shown that the strength, velocity and evolution of the predicted \ion{He}{I} 10830~\AA \ feature is model dependent. 
This shows that the feature is also observer direction dependent, since the 1D models are representative of different directions in the 3D model. 
These models support that a \ion{He}{I} 10830~\AA \ feature is a direct signature of the double detonation, however, the feature can be blended with the \ion{Mg}{II} 10927~\AA \ feature.

\subsection{Ionisation state}
\label{sec:ionisationstate}

\begin{figure}
    \begin{subfigure}[b]{0.45\textwidth}
            \caption{1D models at $\theta=0^\circ$}
            \includegraphics[width=\textwidth]{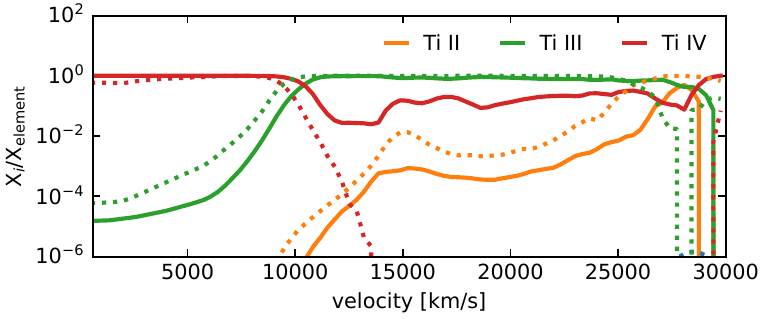}
    \vspace{+0.1cm}
    \end{subfigure}

    \begin{subfigure}[b]{0.45\textwidth}
        \caption{1D models at $\theta=90^\circ$}
        \includegraphics[width=\textwidth]{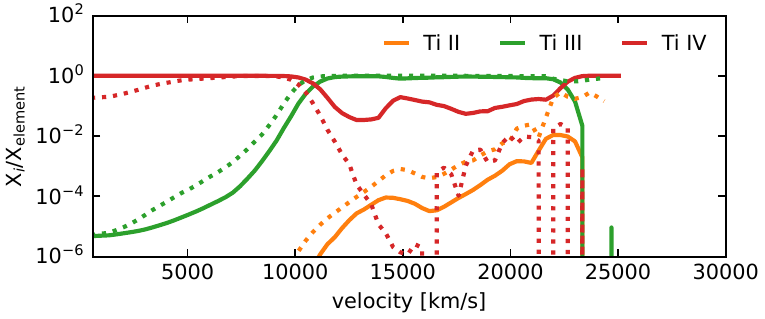}
    \vspace{+0.1cm}
    \end{subfigure}

    \begin{subfigure}[b]{0.45\textwidth}
        \caption{1D models at $\theta=180^\circ$}
        \includegraphics[width=\textwidth]{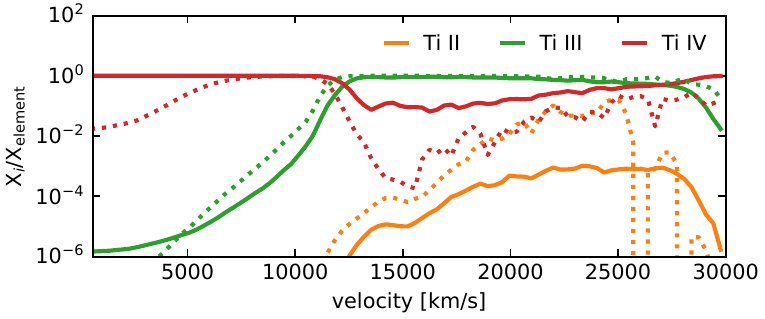}
\vspace{+0.1cm}
    \end{subfigure}

    \caption{Ion fractions of Ti in each of the 1D models at 18 days. The dotted lines show the \textsc{artis-classic} simulations and the solid lines show the \textsc{artis-nlte} simulations.
    For each model the ionisation state is higher in the \textsc{artis-nlte} simulation.
    Note the lower maximum ejecta velocities for the model at $\theta = 90^\circ$ (see Figure~\ref{fig:abundances1D}).
    }
    \label{fig:ionfractions}
\end{figure}

The bluer light curve colours and reduced absorption in the \textsc{artis-nlte} simulations are a result of the higher ionisation state predicted by \textsc{artis-nlte}.
In Figure~\ref{fig:ionfractions} we show the ion fractions of Ti for each simulation.
In each case, \ion{Ti}{III} is generally the dominant species in the outer ejecta. In the \textsc{artis-classic} simulations, \ion{Ti}{II} is generally the next most abundant, however, in the \textsc{artis-nlte} simulations, \ion{Ti}{IV} is the next most abundant.
The fraction of \ion{Ti}{II} is lower in the \textsc{artis-nlte} simulations.
It is this difference in ionisation that leads to the reduced \ion{Ti}{II} absorption in the spectra.
The overall ionisation state of all elements included shows a similar increase, which leads to the bluer colours.

\begin{figure}
\includegraphics[width=0.45\textwidth]{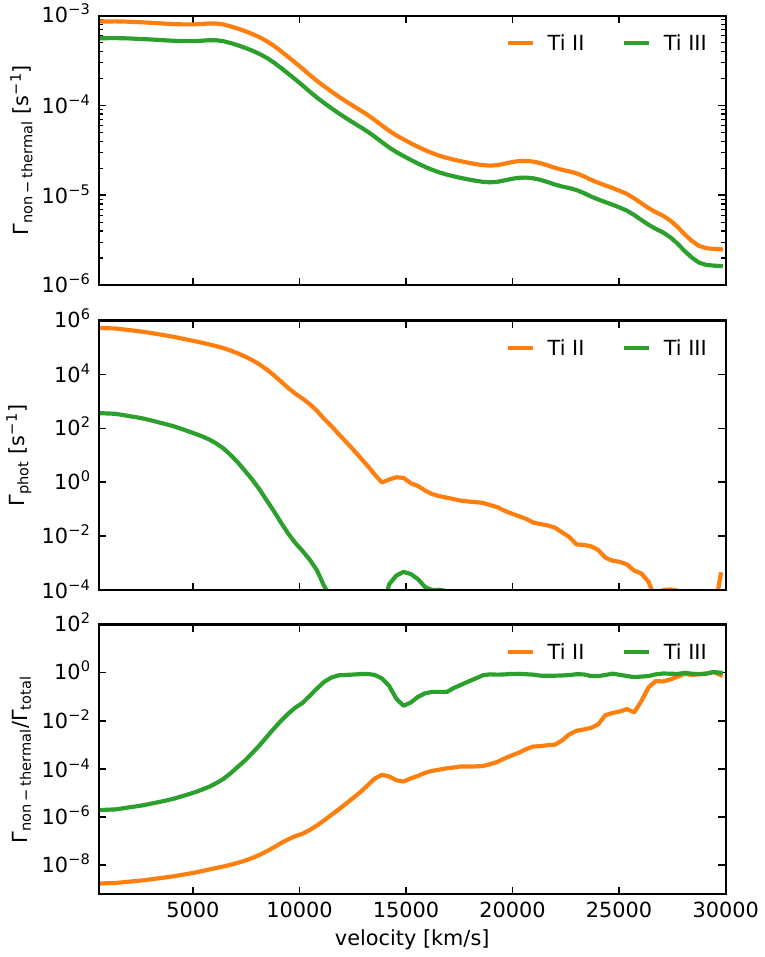}

\caption{Non-thermal ionisation rate coefficient, $\Gamma_{\rm non-thermal}$, (upper panel) and photoionisation rate coefficient, $\Gamma_{\rm phot}$, (middle panel) at 18 days for the \textsc{artis-nlte} model at $\theta = 0^\circ$. Also shown is the ratio of the non-thermal rate coefficient to the sum of the photoionisation and non-thermal rate coefficients, $\Gamma_{\rm total}$ (lower panel).
The 1D models at $\theta = 90^\circ$ and $\theta = 180^\circ$ show similar behaviour.
}
\label{fig:ionisationrates}
\end{figure}

The higher ionisation is predominantly a result of the bluer radiation field predicted by \textsc{artis-nlte}.
As can be seen in Figure~\ref{fig:ionisationrates}, photoionisation is the dominant source of ionisation in the inner ejecta.
In the outer ejecta, non-thermal ionisation also contributes to the higher ionisation, in particular for \ion{Ti}{III} to \ion{Ti}{IV}, however, photoionisation remains the dominant source of ionisation from \ion{Ti}{II} to \ion{Ti}{III} in the He shell ash.

\subsection{Sub-Chandrasekhar mass detonation model with no helium shell detonation}

\begin{figure*}
    \includegraphics[width=0.95\textwidth]{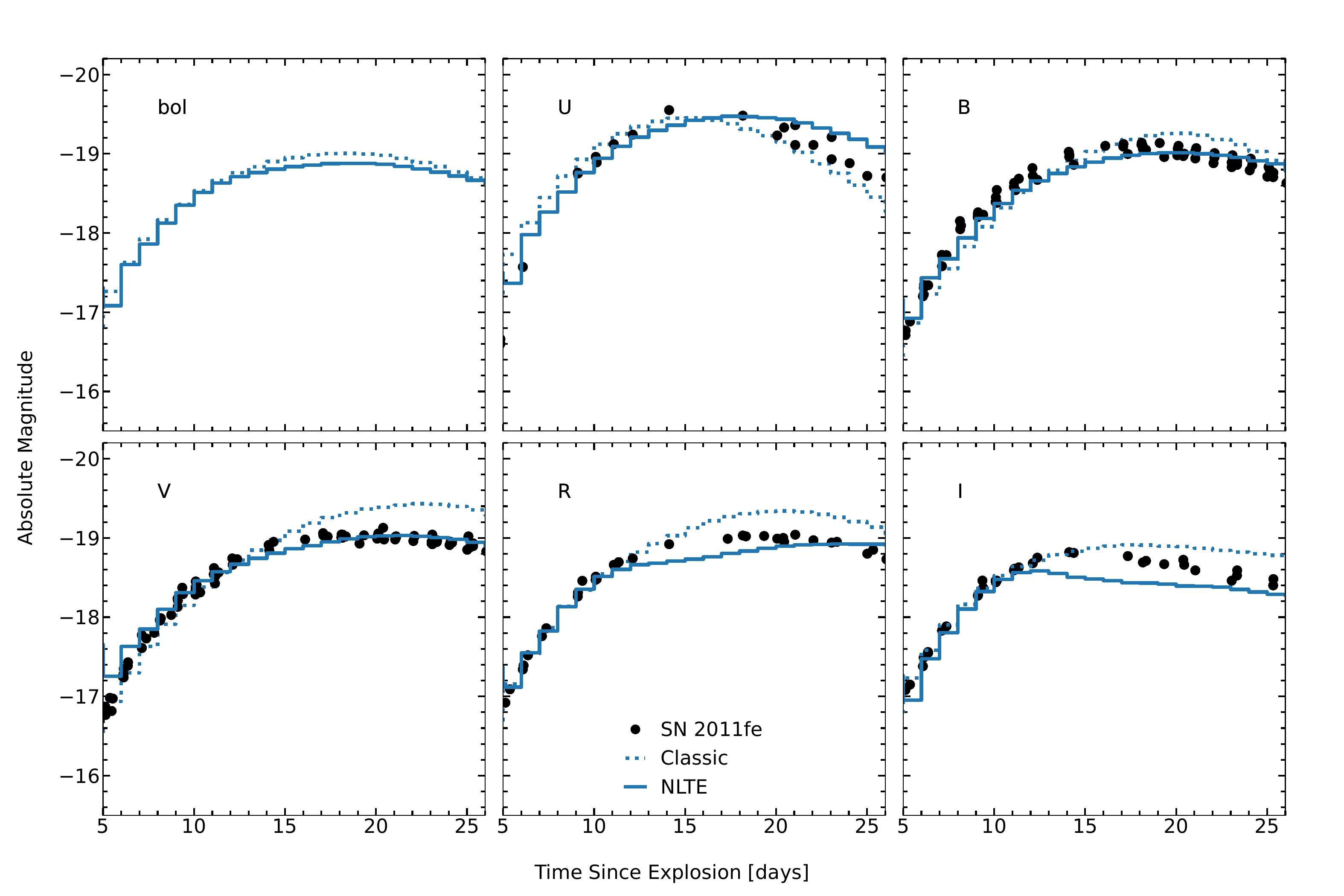}

    \caption{Light curves from an explosion model of a bare C/O sub-M$_{\rm Ch}$ white dwarf with no helium shell detonation, simulated using \textsc{artis-classic} (dotted lines) and \textsc{artis-nlte} (solid lines).}
    \label{fig:lightcurves-puredet}
\end{figure*}

A number of studies have considered the detonation of a bare C/O white dwarf, with no initial helium detonation providing a physical mechanism to ignite the core white dwarf detonation \citep[e.g.,][]{sim2010a, blondin2017a, shen2018b, shen2021a}.
Radiative transfer simulations for these models show good agreement with observations of normal SNe Ia, and they provide motivation for the double detonation scenario.
Here we carry out a non-LTE simulation using \textsc{artis-nlte} for the 1.06 M$_\odot$ C/O bare white dwarf detonation model presented by \citet{sim2010a}, which was ignited artificially and did not include a helium shell.
The light curves from this simulation are shown in Figure~\ref{fig:lightcurves-puredet}, and are compared to a simulation of the same explosion model using \textsc{artis-classic}.
While we do find quantitative differences in the light curves, the differences are not as extreme as for the double detonation models, particularly compared to the models representative of the fainter directions at $\theta = 0^\circ$ and $\theta = 90^\circ$.

\begin{figure*}

    \includegraphics[width=0.45\textwidth]{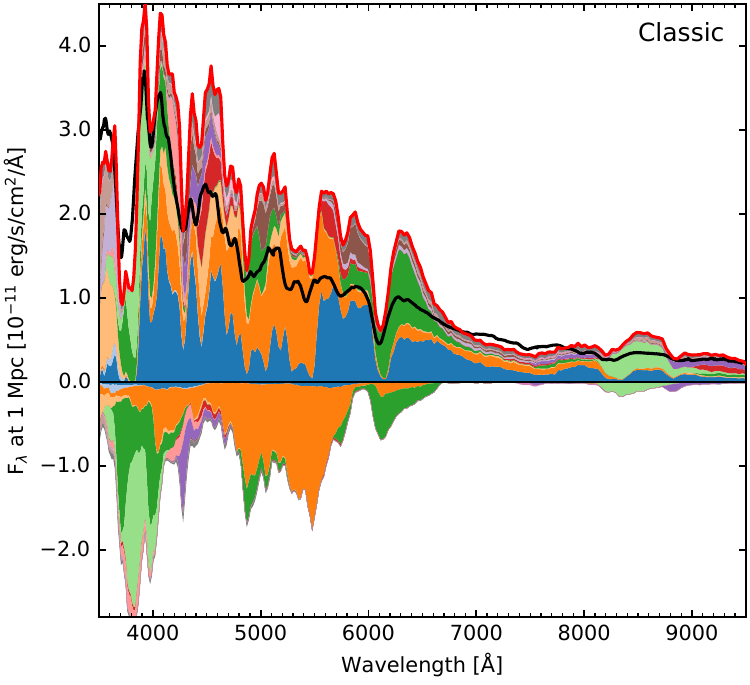}
    \includegraphics[width=0.45\textwidth]{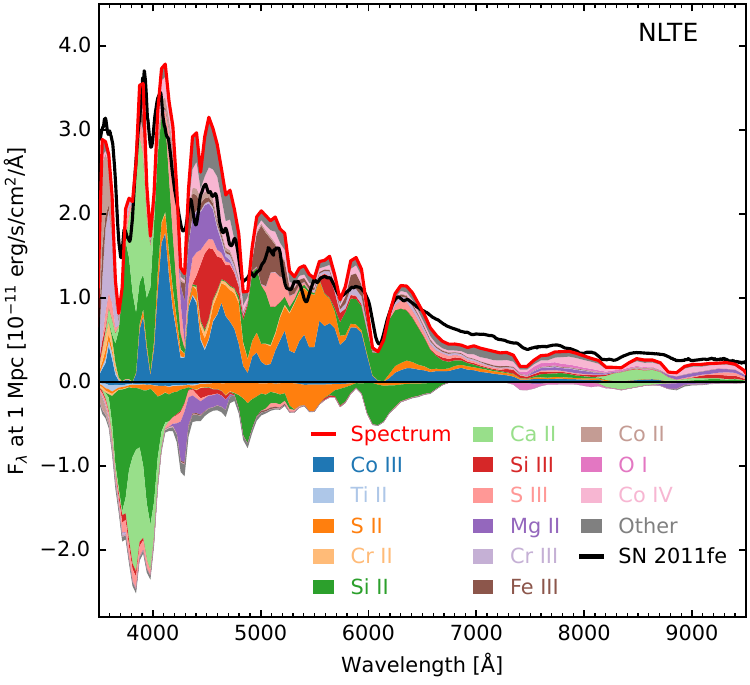}

    \caption{Spectra at 18 days, simulated using \textsc{artis-classic} (left) and \textsc{artis-nlte} (right) for a bare C/O sub-M$_{\rm Ch}$ white dwarf detonation.
    Note that the selected ions shown are the same as in Figure~\ref{fig:specemission_1DdoubledetNLTE} for ease of comparison.
    A near maximum light spectrum of SN~2011fe is plotted for comparison.}
    \label{fig:specemissionpuredet}
\end{figure*}

The contributions to the spectra of the pure detonation model using \textsc{artis-nlte} and \textsc{artis-classic} are shown in Figure~\ref{fig:specemissionpuredet}.
We find the same result that the \textsc{artis-nlte} simulation is more highly ionised than the \textsc{artis-classic} simulation, and find less overall contribution to the spectra from singly ionised elements, in particular \ion{S}{II} and \ion{Ca}{II}.
The \textsc{artis-nlte} simulation leads to better agreement with SN~2011fe.
\citet{shen2021a} found similar behaviour in their non-LTE simulations for pure detonation models with no helium shell, finding that these models were bluer and showed better agreement with normal SNe Ia compared to their LTE simulations.

The difference between the \textsc{artis-classic} and \textsc{artis-nlte} simulations, however, is not as extreme as for the double detonation simulations.
In the double detonation simulations, the absorption at blue wavelengths is dominated by Cr and Ti, which are products of the helium shell detonation.
The pure detonation model does not contain significant amounts of these elements.
This suggests that the impact of the full non-LTE treatment is more critical for the double detonation scenario where heavy elements are produced in the outer ejecta from the helium detonation.
In future, this should be tested for a broader range of SNe~Ia explosion models to further understand the impacts of non-LTE effects.

\section{Conclusions}

We have presented synthetic observables for double detonation
simulations calculated in full non-LTE using \textsc{artis-nlte}.
We aimed to investigate
the effects of a full non-LTE treatment on the predictions of the
synthetic observables for the double detonation scenario.
Previous simulations without a full non-LTE treatment found that the products
of the helium shell detonation caused strong absorption features, leading
to colours significantly redder than observed in normal SNe~Ia.
We found that \textsc{artis-nlte} predicted light curves showing
colours significantly less red than those predicted by
\textsc{artis-classic} and show similar colours to those of normal SNe Ia.
In particular, the U and B band light curves are significantly affected by the more detailed treatment of ionisation.
All models became bluer, in particular in U-B and B-V colours.

We constructed
1D models representative of lines of sight in a 3D double detonation
explosion model to indicate the potential direction dependence.
\textsc{artis-classic} double detonation simulations showed too large a viewing angle dependence compared to observations \citep{collins2022b}. The viewing angle dependence indicated by the 1D models at early phases is significantly reduced
in the \textsc{artis-nlte} simulations compared to the \textsc{artis-classic} simulations, improving agreement with observations.
This should be investigated further in future using multi-dimensional simulations.

The \textsc{artis-nlte} simulations showed a better
match to many features of the normal SN~2011fe than the \textsc{artis-classic} simulations.
In the \textsc{artis-nlte} simulations, the ejecta are more highly ionised compared to \textsc{artis-classic}
simulations, and it is predominantly the reduction in contributions from 
singly ionised heavy elements produced in the helium detonation (e.g. \ion{Ti}{II}, \ion{Cr}{II})
responsible for the changes found in the light
curves, colours and spectra.
This shows that the model colours are extremely sensitive to the ionisation
state in the case of double detonation simulations.

For comparison, we also presented \textsc{artis-nlte} simulations for an explosion model of a bare C/O white dwarf, which did not include a helium shell.
While noticing an improved agreement with observations, we showed that the difference between the \textsc{artis-nlte} and \textsc{artis-classic} simulations were not as large for this explosion model, highlighting that the majority of the difference in the double detonation models comes from the improved ionisation treatment of the
helium shell detonation products.

The non-LTE modelling presented here more reliably reflects the true physics than is represented by the approximations made in \textsc{artis-classic}, or in LTE. 
However, this comes with considerably more computational cost to simulate in non-LTE.
\textsc{artis} is a multi-dimensional code, and we aim in future to be able to carry out multi-dimensional non-LTE simulations, which would verify the extent to which our 1D models can represent viewing angle effects in a multi-dimensional non-LTE simulation.
Additionally, since double detonation models show significant variation with initial white dwarf and helium shell masses, more double detonation models should be investigated in non-LTE in future.

\section*{Acknowledgements}

CEC is grateful for support from the Department for the Economy (DfE).
SAS and FPC, acknowledge funding from STFC grant ST/X00094X/1.
This project has received funding from the European Union’s Horizon Europe
research and innovation programme under the Marie Skłodowska-Curie grant
agreement No. 101152610.
This work was partly funded by the European Union (ERC, HEAVYMETAL, 101071865). FKR acknowledges funding by the European Union (ERC, ExCEED, project number 101096243). Views and opinions expressed are however those of the authors only and do not necessarily reflect those of the European Union or the European Research Council. Neither the European Union nor the granting authority can be held responsible for them.
LJS acknowledges support by the European Research Council (ERC) under
the European Union's Horizon 2020 research and innovation
program (ERC Advanced Grant KILONOVA No. 885281) and support by Deutsche Forschungsgemeinschaft (DFG, German Research Foundation) - Project-ID 279384907 - SFB 1245 and MA 4248/3-1.
The work of FKR is supported by the Klaus Tschira Foundation and by the Deutsche Forschungsgemeinschaft (DFG, German Research Foundation) -- RO 3676/7-1, project number 537700965.
The authors gratefully acknowledge the Gauss Centre for Supercomputing e.V.
(www.gauss-centre.eu) for funding this project by providing computing time
through the John von Neumann Institute for Computing (NIC) on the GCS
Supercomputer JUWELS at J\"ulich Supercomputing Centre (JSC).
This work used the DiRAC Data Intensive service (CSD3) at the University of Cambridge, managed by the University of Cambridge University Information Services on behalf of the STFC DiRAC HPC Facility (www.dirac.ac.uk). The DiRAC component of CSD3 at Cambridge was funded by BEIS, UKRI and STFC capital funding and STFC operations grants. This work also used the DiRAC Memory Intensive service (Cosma8) at Durham University, managed by the Institute for Computational Cosmology on behalf of the STFC DiRAC HPC Facility (www.dirac.ac.uk). The DiRAC service at Durham was funded by BEIS, UKRI and STFC capital funding, Durham University and STFC operations grants. DiRAC is part of the UKRI Digital Research Infrastructure. The authors are grateful for computational support by the VIRGO cluster at GSI.
\href{https://github.com/artis-mcrt/artis}{\textsc{artis}}\footnote{\href{https://github.com/artis-mcrt/artis/}{https://github.com/artis-mcrt/artis/}}
\citep{sim_artis} was used to carry out the radiative transfer simulations.
\href{https://github.com/artis-mcrt/artistools}{\textsc{artistools}}\footnote{\href{https://github.com/artis-mcrt/artistools/}{https://github.com/artis-mcrt/artistools/}}
\citep{artistools}
were used for data processing and plotting.

\section*{Data Availability}

The data underlying this article will be shared on reasonable request to the corresponding author.
 



\bibliographystyle{mnras}
\bibliography{astrofritz} 




\appendix

\section{1D models created from a 3D model}
\label{sec:verify1Dmodels}

\subsection{Constructing the 1D models}

To create 1D models representative of a given line of sight, 
we construct cones from the 3D model around a line of sight,
and average these onto a single line to create a 1D profile.
Each cone is defined such that the angle 
between the line of sight and the cone boundary is 15$^\circ$ (where the full opening
angle of the cone is 30$^\circ$).
The abundance and density profiles within the cone are averaged 
along the line of sight.
The 1D model at $\theta = 0^\circ$ is constructed from a cone around the
positive $z$-axis.
The 1D profile at $\theta = 90^\circ$ was constructed from a cone
around the negative $y$-axis.
We do not expect the choice of this direction to be significantly different
from other equatorial lines of sight, as the ejecta are close to
symmetric in the azimuthal direction.
The model at $\theta = 180^\circ$ was constructed from a cone around the
negative $z$-axis.
The 1D profiles in the positive $z$, negative $y$ and negative $z$
directions are similar to $\theta = 0^\circ$, $\theta = 90^\circ$
and $\theta = 180^\circ$ calculated in the 3D
simulations presented by \citet{gronow2020a}.
We therefore refer to each of these 1D models by angle of the line of sight in the 3D model.

\begin{figure*}
\includegraphics[width=0.95\textwidth]{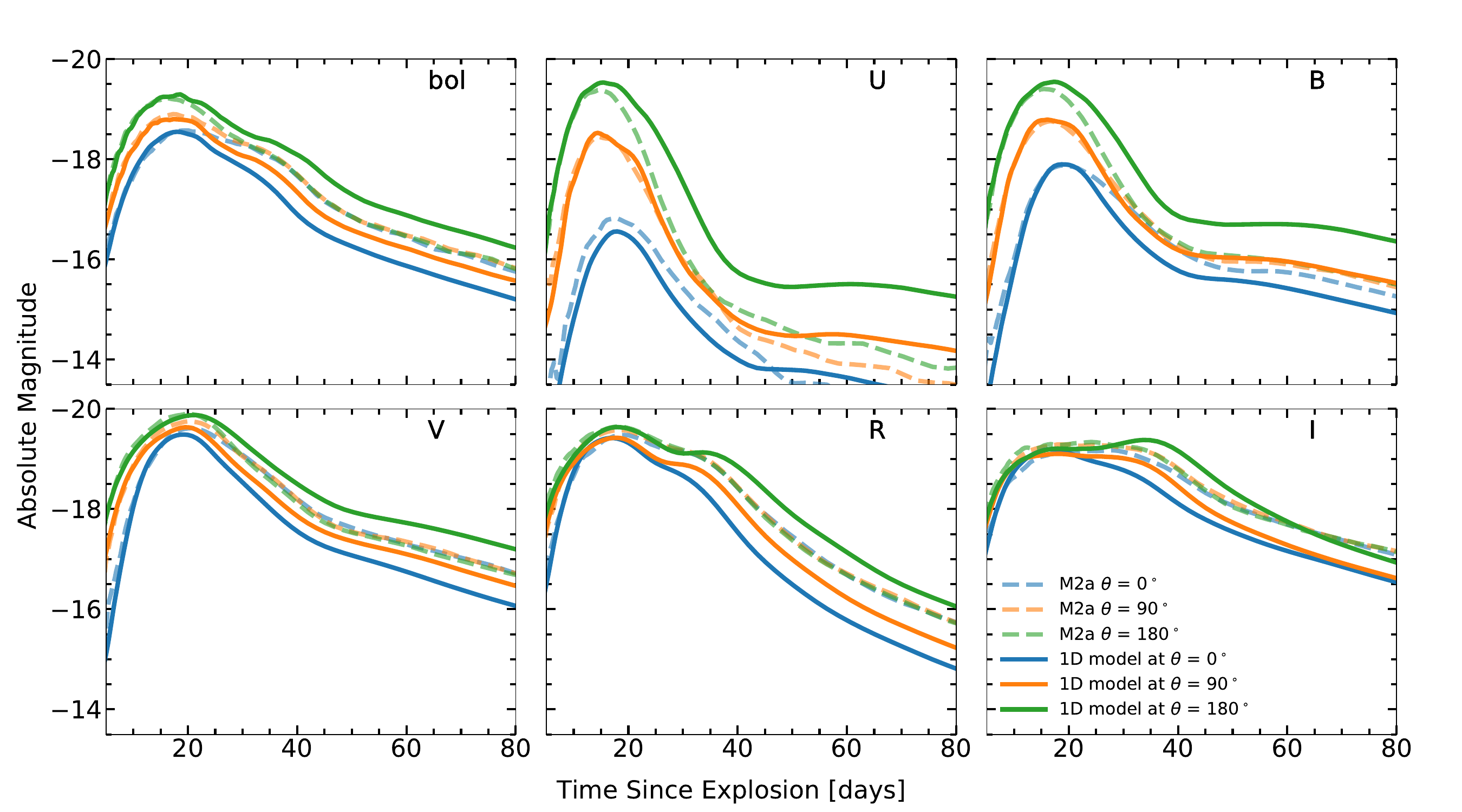}

\caption{Light curves for 1D models constructed from the 3D Model M2a (solid lines). These are compared to the line of sight dependent light curves for the 3D simulation of Model M2a.
\textsc{artis-classic} was used for each of these simulations.
}
\label{fig:lightcurves1Dcomparedto3D}
\end{figure*}

\begin{figure}
\begin{center}
\includegraphics[width=0.4\textwidth]{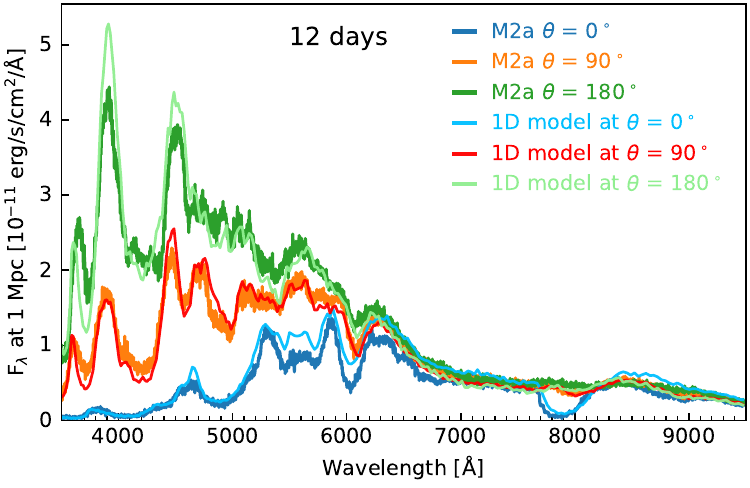}

\includegraphics[width=0.4\textwidth]{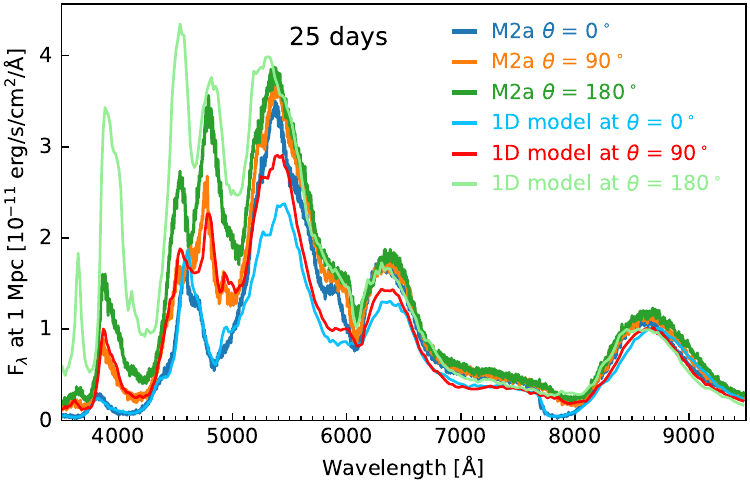}
\end{center}

\caption{Spectra at 12 and 25 days showing the comparison between the line of sight spectra in Model M2a, and the 1D models representative of these viewing angles.
Around 12 days the 1D model spectra are able to reproduce the viewing angle effects of the 3D model, however by 25 days the 1D models over-predict the viewing angle effects.}
\label{fig:spectracompare1Dto3D}
\end{figure}

To test how well the 1D profiles are able to match the 3D lines of sight
we use \textsc{artis-classic} to calculate light curves and spectra
for the 1D profiles so that we can directly compare these calculations
to the 3D results, which were also calculated using \textsc{artis-classic}.

\subsubsection{1D light curves compared to 3D}
We compare the light curves from both the 1D simulations and the 
3D simulation in Figure \ref{fig:lightcurves1Dcomparedto3D}.
We find that the 1D profiles are a good match to the viewing angle 
light curves from the 3D simulation during the initial rise of the 
light curves. 
Around maximum, while they are not able to exactly reproduce the 
3D simulation, they capture the general trends observed for these 
lines of sight, and in comparison to the scale of apparent discrepancies 
with observations discussed by \citet{gronow2020a}, 
the scale of the differences around maximum are small.

In the 3D simulation we find that after $\sim$ 30 days 
(earlier in redder bands) the differences between the viewing angles in the 3D simulation
become small, such that we no longer find strong viewing angle effects.
The 1D profiles do not reproduce this trend. 
The strong viewing angle effects remain at all times for these.
Therefore the 1D profiles significantly overestimate the viewing 
angle effects at later times.
As the ejecta are optically thick at early times, 
most of the emission comes from the ejecta in the line of sight. 
This is why we see agreement between the 3D simulation and the 
1D simulations at these times.
However, over time as the ejecta expand and become more optically thin, 
the emission in a given line of sight increasingly includes contributions from 
regions of the ejecta that are not represented in the 1D profiles, 
and in turn, not all photons in a line of sight will be emitted in 
that direction.
We can therefore use the 1D profiles to investigate viewing angle effects
at early times and around maximum, however at later times the 1D profiles 
do not show good agreement with the multi-dimensional light curves.

Recently, \citet{boos2024a} showed a comparison between 1D models made from wedges in different observer directions from a 2D double detonation explosion model.
In their 1D simulations, additional energy was injected into the simulation to match the bolometric luminosity to that in the 2D model in the corresponding line of sight, resulting in better apparent agreement between the 1D and multi-dimensional simulations until later times.
We do not include any such scaling of the bolometric luminosity. 
Since the luminosities here are not scaled, this allows us to investigate the impact of non-LTE on the bolometric luminosity.

\subsubsection{1D spectra compared to 3D}

As with the light curves, we find that the spectra from the 1D simulations 
are a good match to the 3D simulation at early times, 
but at later times the viewing angle effects are over estimated by the 
1D simulations.
This can be seen in Figure~\ref{fig:spectracompare1Dto3D},
where we compare the 1D simulations to the line of sight spectra from 
the 3D simulation.
Overall, the spectra from the 1D simulations at 12 days after explosion are a good match
to those of the 3D simulation, and therefore at these times we can use 
the 1D profiles to investigate the 3D viewing angle spectral effects. 
By 25 days we can see from Figure~\ref{fig:spectracompare1Dto3D} that
although generally the 1D simulations are able to produce the same 
features as the 3D simulation, the strengths of the features are not 
well matched.

This shows that the 1D profiles constructed from a given line of sight in a 3D model are able to represent the results
of 3D simulations at early times, before maximum light.
We find that around maximum light the 1D profiles still show reasonably
good agreement with the 3D lines of sight, however, after $\sim 30$ days
from the time of explosion, the viewing angle effects represented by the
1D profiles significantly over-estimate the level of the viewing angle
effects found in the 3D model.
We therefore use these 1D profiles as an initial test case for \textsc{artis-nlte}
to investigate the effects at early times on the double detonation explosion scenario.


\bsp	
\label{lastpage}
\end{document}